\documentclass[11pt,ukenglish]{amsart}
\usepackage{amssymb}
\usepackage{graphicx}
\usepackage{amscd}

\theoremstyle{plain}

\newtheorem{definition}{Definition}

\newtheorem{proposition}{Proposition}

\numberwithin{equation}{section}

\input{tcilatex}

\begin{document}
\title[Nonlinear Hamiltonian Kinetic Equations]{Uniformization of Nonlinear
Hamiltonian Systems of Vlasov and Hartree Type}
\author{V.P. Belavkin and V.P. Maslov}
\address{Moscow Institute of Electronics and Mathematics\\
Moscow 109028, USSR.}
\email{vpb@maths.nott.ac.uk}
\urladdr{http://www.maths.nott.ac.uk/personal/vpb/}
\date{23 February 1977}
\thanks{This paper is translated by Plenum Publishing Corporation from
Teoreticheskaya i matematicheskaya Fizika, Vol. 33 (1977), No. 1, pp17--31}

\begin{abstract}
Nonlinear Hamiltonian systems describing the abstract Vlasov and Hartree
equations are considered in the framework of algebraic Poissonian theory.
The concept of uniformization is introduced; it generalizes the method of
second quantization of classical systems to arbitrary Hamiltonian
(Lie-Jordan, or Poissonian) algebras, in particular to the algebras of
operators in indefinite spaces. A functional calculus is developed for the
uniformized observables, which unifies the calculi of generating functionals
for classical, Bosonian and Fermionian multiparticle variables. The
nonlinear kinetic equation of Vlasov and Hartree type are derived for both
classical and quantum multiparticle Hamiltonian systems in the mean field
approximation. The possibilities for finding approximate solutions of these
equations from the solutions of uniformized linear equations are
investigated.
\end{abstract}

\maketitle

\section{Introduction}

Recently, much interest \cite{1} has been devoted to studying the region of
the spectrum of the Hamiltonians of certain self-interacting boson fields
corresponding to particle-like (soliton) solutions of the classical mean
field non-linear equations. Such excitations can be naturally interpreted as
composite stable quantum particles that have finite energy even in the
classical limit. As a comparison \cite{2} of the results for exactly
solvable models showed, there is a direct correspondence between the quantum 
\cite{3} and classical \cite{4} particle-like solutions. Frequently, it is
simplest to obtain such solutions directly for the quantum case, and not for
the corresponding classical case, for which they were obtained only very
recently by the solution of the inverse scattering problem.

This correspondence can be viewed as a heuristic basis of an asymptotic
method for finding soliton solutions of nonlinear equations in not only
classical field theory but also for the difference equations that describe
the dynamics of nonlinear crystals, and also certain nonlinear equations of
quantum electrodynamics. This method, which we call the uniformization
method, can be regarded mathematically as a new method of linearization
leading to asymptotically exact solutions of nonlinear equations in the
large number limit by the infinite increase in the number of degrees of
freedom.

Applied to ordinary nonlinear Schr\"{o}dinger equations, the uniformization
method essentially reduces to second quantization and the construction of a
quasiclassical soliton solution from the solution of the $n$-particle
equation for sufficiently large $n$. However, this interpretation no longer
holds for the spaces with indefinite metric. Applying this method to the
nonlinear equation of quantum electrodynamics (\cite{5}, p.11) in which the
local time enters as a fourth coordinate, we arrive at a many-time formalism
that reduces the solution of these nonlinear equations to the solution of
ultrahyperbolic equations that may not have independent physical meaning.

Therefore, we set forth the uniformization method on an abstract algebraic
basis without referring to its physical interpretation which is contained in
the concrete functional realizations. As the object of uniformization, we
choose the abstract Vlasov equation, which generalizes the classical and
quantized Vlasov equations considered in \cite{5}, and also equations that
reduce to equations of Hartree type in spaces with indefinite metric.

For the classical case, the uniformization method is essentially the
transition from a field system described by the Vlasov kinetic equation to a
system with variable number of particles, i.e., it is the procedure which is
the inverse of the derivation of the Vlasov equation from the Bogolyubov
chain of $n$-particle equations \cite{6}. In the quantum case, the
uniformization method gives a unified algebraic basis for the method of
second quantization of bosons and fermions with a superselection rule that
avoids the consideration of unphysical - odd - observables that do not
commute with the particle number operator. The developed calculus of
representing functionals for such even operators combines and replaces the
calculus of symbols of commuting variables (Bargmann \cite{6}) and
anticommuting variables (Berezin \cite{7}). In such an approach, the
difference between bosons and fermions does not reside in the operators but
only in the multiparticle tensor-states.

In the general case it is shown that the algebra of polynomial observables
of the uniformized system is isomorphic to the direct sum of the tensor
powers of the Hamiltonian algebra generated by the elementary observables of
the uniformized system. In the case of algebras of operators in a space with
indefinite metric, such a representation generalizes the Fock to a
pseudo-Fock representation.

First, we set forth the basic concepts and the definitions relating to
nonlinear Hamiltonian systems of Vlasov and Hartree type, and also the
uniformization method, showing that such nonlinear kinetic equations
naturally appear as characteristic equations in the mean-field approximation
of the uniformized Hamiltonian infinite systems. Then we consider the
possibility of finding approximate solutions of these nonlinear equations,
in particular approximate soliton solutions from the corresponding solutions
of the uniformized equations.

\section{Abstract Vlasov and Hartree Equations}

\subsection{ \ }

We consider a dynamical system whose states are described by the elements $%
\rho $ of a real linear space $\mathcal{L}$ as the predual space of linear
functionals $A\mapsto \left\langle \rho ,A\right\rangle $ on an abstract
Hamiltonian algebra $\mathcal{A}$ with respect to some bilinear form $%
\left\langle \rho ,A\right\rangle $ having real values for all $\rho \in 
\mathcal{L}$ and $A=A^{\star }\in \mathcal{A}$\footnote{%
One may assume that $\mathcal{L}$ is a Fr\'{e}chet space with the
complexified dual of all linear continuous functionals $\rho \mapsto
\left\langle A,\rho \right\rangle $ given by a $\star $-algebra $\mathcal{A}$%
.}. By the Hamiltonian algebra we mean any complex-linear unital $\star $%
-algebra $\mathcal{A}$ in which besides the \textquotedblleft
ordinary\textquotedblright , not necessarily commutative and associative,
product $A\cdot A^{\star }\in \mathcal{A}$, there is also defined the
Poisson brackets $\left\{ H,A\right\} $ as a bilinear antisymmetric map $%
\mathcal{A}\times \mathcal{A}\rightarrow \mathcal{A}$ with the property of
continuous derivation%
\begin{equation*}
\left\{ H,A\cdot A^{\star }\right\} =A\cdot \left\{ H,A^{\star }\right\}
+\left\{ H,A\right\} \cdot A^{\star }
\end{equation*}%
having anti-Hermitian values $\left\{ A^{\star },A\right\} ^{\star
}=-\left\{ A^{\star },A\right\} $ in contrast to $\left( A\cdot A^{\star
}\right) ^{\star }=A\cdot A^{\star }$ (for more details see \cite{7}). If
the ordinary product is associative, the Hamiltonian algebra is said to be
Poissonian. In the commutative case $A\cdot A^{\star }=A^{\star }\cdot A$
the real part $\Re \mathcal{A}=\left\{ A=A^{\star }\right\} $ of the
Hamiltonian algebra is a real Hamiltonian subalgebra such that $\mathcal{A}$
can be considered as a complexification of $\mathcal{R}=\Re \mathcal{A}$
which is usually taken to be a real Lie-Jordan algebra \cite{8}. If the
algebra $\mathcal{A}$ is both commutative and associative with respect to
the product $A\cdot B$ like the algebra of smooth scalar-valued functions on
a Hamiltonian phase space, it is called classical Hamiltonian, or,
equivalently Poissonian algebra.

We give the name Vlasov Hamiltonian to every real element $H=H^{\star }$ $%
\in \mathcal{A}$ that may depend continuously on $\rho \in \mathcal{L}$ as
the derivative $\delta \gamma (\rho )=H(\rho )$ of some, in general
nonlinear $\mathcal{A}$-smooth functional $\gamma :\rho \mapsto \gamma (\rho
)$, called Hamiltonian functional on $\mathcal{L}$ ($\gamma $ has real
number values $\gamma \left( \rho \right) \in \mathbb{R}$). By $\mathcal{A}$%
-smooth we mean such smooth functional which has the derivative as a
continuous functional $\rho \mapsto \delta \gamma \left( \rho \right) $
defined for each $\rho \in \mathcal{L}$ as an element $H\left( \rho \right) $
of $\mathcal{A}$ such that 
\begin{equation}
\left\langle \sigma ,\delta \right\rangle \gamma \left( \rho \right) :=\frac{%
\partial \gamma (\rho +t\sigma )}{\partial t}\mid _{t=0}=\left\langle \sigma
,H(\rho )\right\rangle ,\qquad \forall \sigma \in \mathcal{L}.  \label{A}
\end{equation}

Let us assume that the linear operator $\rho \mapsto \left\{ \rho ,H\right\} 
$ is well-defined for each value $H=H\left( \rho \right) $ of a Vlasov
Hamiltonian $H$ as the predual to the derivation $A\mapsto \left\{
H,A\right\} $ on the algebra $\mathcal{A}$ such that\footnote{%
To this end one may assume that the derivation $A\mapsto \left\{ H,A\right\} 
$, as well as the action $A\mapsto H\cdot A$ are $\sigma \left( \mathcal{A},%
\mathcal{L}\right) $-continuous for each $H\in \mathcal{A}$.} 
\begin{equation}
\left\langle \left\{ \rho ,H\right\} ,A\right\rangle =\left\langle \rho
,\left\{ H,A\right\} \right\rangle ,\qquad \forall \rho \in \mathcal{L},A\in 
\mathcal{A}.  \label{B}
\end{equation}%
We give the name \textquotedblleft abstract Vlasov
equation\textquotedblright\ to the evolution equation in the state space $%
\mathcal{L}\ni \rho (t)$ of the form 
\begin{equation}
\frac{\partial \rho (t)}{\partial t}=\left\{ \rho (t),H(t,\rho (t))\right\} .
\label{C}
\end{equation}%
where $H(t,\rho )$ is the derivative $\delta \gamma (t,\rho )$ of a
nonlinear time-dependent $\mathcal{A}$-smooth functional $\gamma (t,\rho )$
which may, in general, depend on time $t$.

\begin{proposition}
\label{P1}Equation (\ref{C}) is a classical Hamiltonian system with respect
to the Hamiltonian functional $\gamma (t,\rho )$ which has at least one
linear integral of motion $\nu (\rho )=\left\langle \rho ,I\right\rangle $
generated by the identity $I\in \mathcal{A}$.
\end{proposition}

Indeed, for any differentiable functional $\alpha :\rho \mapsto \alpha (\rho
)$ that does not depend explicitly on $t$ we have in accordance with (\ref{C}%
) 
\begin{equation}
\frac{\partial \alpha (\rho (t))}{\partial t}=\left\langle \left\{ \rho
(t),H(t,\rho (t))\right\} ,\delta \alpha (\rho (t))\right\rangle .  \label{D}
\end{equation}%
Using the definitions (\ref{B}) of the dual operator $A\mapsto \left\{
H,A\right\} $ and remembering that $H=\delta \gamma (\rho )$, we can write (%
\ref{D}) in the form $\frac{\partial \alpha }{\partial t}=\left\{ \gamma
,\alpha \right\} _{cl}$, where 
\begin{equation}
\left\{ \gamma ,\alpha \right\} _{cl}(\rho )=\left\langle \rho ,\left\{
\delta \gamma (\rho ),\delta \alpha (\rho )\right\} \right\rangle  \label{E}
\end{equation}%
is a classical Poisson bracket with respect to the pointwise product $\alpha
\beta $ since%
\begin{equation*}
\left\{ \gamma ,\alpha \beta \right\} _{cl}(\rho )=\alpha \left( \rho
\right) \left\{ \gamma ,\beta \right\} _{cl}(\rho )+\left\{ \gamma ,\alpha
\right\} _{cl}(\rho )\beta \left( \rho \right) .
\end{equation*}%
This means that the algebra $\mathfrak{A}_{cl}$ of smooth functionals $%
\alpha :\mathcal{L}\rightarrow \mathbb{C}$ is a classical Hamiltonian
algebra with commutative and associative Jordan product $\left( \alpha \cdot
\beta \right) (\rho )=\alpha (\rho )\beta (\rho )$ and Lie product (\ref{E})
which generates on the manifold $\mathcal{L}$ a symplectic structure with
respect to which (\ref{C}) is simply the Hamiltonian equation with the
Hamilton functional $\gamma (t,\rho )$. Substituting $\nu (\rho
)=\left\langle \rho ,I\right\rangle $ as $\alpha (\rho )$ in (\ref{E}) and
remembering that in accordance with the derivation property we have $\left\{
H,I\right\} =0$ for any $H$ $\in \mathcal{A}$, we find that $\nu (\rho )$ is
an integral for (\ref{C}): $\left\{ \gamma ,\nu \right\} _{cl}=0$.

Note that the elements $\alpha $ of the algebra $\mathfrak{A}_{cl}$ can be
interpreted as nice (smooth) observables of the described classical system,
taking real or complex values $\alpha (\rho )$ in the states $\rho \in 
\mathcal{L}$, which should be regarded as deterministic point-states
corresponding to zero variances $\left\vert \alpha \right\vert ^{2}\left(
\rho \right) -\left\vert \alpha \left( \rho \right) \right\vert ^{2}=0$
simultaneously for all $\alpha \in \mathfrak{A}_{cl}$. The Hamiltonian
algebra $\mathcal{A}$, whose elements generate \textquotedblleft elementary
classical observables\textquotedblright\ - the linear functional observables 
$\alpha _{A}=\left\langle \rho ,A\right\rangle $ - is here not assumed to be
classical.

\subsection{ \ }

Actually to the Vlasov equation itself there corresponds only the case when $%
\mathcal{A}$ is the classical Poissonian algebra of smooth complex functions 
$A\left( \mathbf{q,p}\right) $ on the single-particle phase space $\mathbb{R}%
^{2d}\ni \left( \mathbf{q,p}\right) $ with ordinary pointwise multiplication
and involution given by complex conjugation \thinspace $A^{\star }\left( 
\mathbf{q,p}\right) =\overline{A\left( \mathbf{q,p}\right) }$, and the
classical Poisson brackets $\left\{ A^{\star },A\right\} =\mathrm{i}\left[
A^{\star },A\right] _{cl}$, where%
\begin{equation*}
\left[ A^{\star },A\right] _{cl}\left( \mathbf{q,p}\right) =\mathrm{i}\left(
\partial _{\mathbf{q}}A^{\star }\cdot \partial _{\mathbf{p}}A-\partial _{%
\mathbf{q}}A\cdot \partial _{\mathbf{p}}A^{\star }\right) \left( \mathbf{q,p}%
\right)
\end{equation*}%
is the classical commutator, i.e. the Poisson bracket divided by $\mathrm{i}=%
\sqrt{-1}$. The space $\mathcal{L}$ of real density functions $\rho \left( 
\mathbf{q,p}\right) $ absolutely integrable in the product with any $A\in 
\mathcal{A}$ is dual to $\mathcal{A}$ with respect to the phase space
integral%
\begin{equation*}
\left\langle \rho ,A\right\rangle =\int \int \rho \left( \mathbf{q,p}\right)
A\left( \mathbf{q,p}\right) \mathrm{d}\mathbf{q}\mathrm{d}\mathbf{p}.
\end{equation*}%
Thus the classical Vlasov equation (\ref{C}) can be regarded as the
Liouville equation with Hamilton function $H(\mathbf{p},\mathbf{q},t,\rho )$
that in general continuously depends on the density function $\rho $
(actually in the original Vlasov equation $H(\mathbf{p},\mathbf{q},t,\rho )$
is an affine function of $\rho $ corresponding to a quadratic form of the
functional $\rho \mapsto \gamma \left( t,\rho \right) $).

One can obtain a different class of the classical Hamiltonian systems of the
form (\ref{C}) by considering the quantized Vlasov equations as in \cite{5},
or even a more general equations when $\mathcal{A}$ is an operator algebra
in a complex-linear Hilbert or pre-Hilbert space $\mathcal{H}$ with scalar
product $\left( \cdot |\cdot \right) $, taking $\mathcal{L}$ as the space of
density operators $\rho $ dual to $\mathcal{A}$ with respect to the
trace-pairing $\left\langle \rho ,A\right\rangle =\mathrm{Tr}(\rho A)$. The
Poisson bracket with respect to the noncommutative operator product in $%
\mathcal{A}$ can be defined by the usual commutator $\left[ A^{\star },A%
\right] =A^{\star }A-AA^{\star }$ as $\left\{ A^{\star },A\right\} =\mathrm{i%
}\left[ A^{\star },A\right] $, so that (\ref{C}) takes the form of a
quantized Liouville, or von Neumann equation%
\begin{equation*}
\frac{\partial \rho (t)}{\partial t}=\mathrm{i}\left[ \rho (t),H(t,\rho (t))%
\right] ,
\end{equation*}%
with $\rho $-dependent Hamilton operator $H(t,\rho )$. Note that the real
part $\Re \mathcal{A}$ of the nocommutative operator algebra $\mathcal{A}$
is not invariant under the associative operator product $AB$, but it is a
nonassociative real Hamiltonian algebra with respect to the symmetrized
operator product $A\cdot B=\frac{1}{2}(AB+BA)$.

One can also have intermediate semiquantized cases obtained, for example, by
partial quantization of the Vlasov equation. In all these cases, if the
states $\rho $ satisfy the positivity condition $\left\langle \rho ,A^{\star
}A\right\rangle \geqslant 0$ and the normalization condition $\left\langle
\rho ,I\right\rangle =1$ they can have a probability interpretation by
regarding the forms $\left\langle \rho ,A\right\rangle $ not as exact values
of elementary observables but as expected values of the results of
individual measurements of these observables. However this probabilistic
interpretation cannot be extended onto the nonlinear kinetic equation (\ref%
{C}) even though they preserve the positivity of $\rho $ and its
normalization, since they do not preserve statistical mixtures except in the
linear case. As we shall see, the the nonlinear kinetic equations describe
not the individual but a collective dynamics of the infinite ensemble of the
corresponding classical or quantum particles in the mean field approximation.

Finally, our treatment also encompasses the (general) case when $\mathcal{A}$
is an operator $\star $-algebra in a space $\mathcal{H}$ with indefinite
metric, i.e., when $\left( \cdot |\cdot \right) $ is an arbitrary
nondegenerate Hermitian form (indefinite scalar product). In this case, the
set of operators of the form $AA^{\star }$ does not form a cone, and the
states $\rho $ cannot have a probability interpretation of the individual
particles. They can only be regarded as pure states of the corresponding
classical field system, or as a quasi-classical limit of an infinite quantum
field system. Numerous examples of such a situation arise in a
quasiclassical treatment of boson and fermion fields with indefinite
commutation relations in \cite{9}.

Suppose $\psi \mapsto \psi ^{\ast }$ is a canonical mapping of the space $%
\mathcal{H}$ into the dual space $\mathcal{H}^{\ast }$ defined by the
equation $\psi ^{\ast }\varphi =\left( \varphi |\psi \right) ,\forall
\varphi \in \mathcal{H}$, and $\psi \psi ^{\ast }$ are one-dimensional
operators in $\mathcal{H}$ acting in accordance with the formula $\psi \psi
^{\ast }\varphi =\left( \varphi |\psi \right) \psi $. The space $\mathcal{L}$
generated by the operators $\psi \psi ^{\ast }$ is the dual to the algebra
of self-adjoint operators in $\mathcal{H}$ with respect to the bilinear form 
$\left\langle \rho ,A\right\rangle $ generated by the form $\left( A\psi
|\psi \right) $ on the elements $\rho =\psi \psi ^{\ast }$. We consider (\ref%
{C}) with the commutator Poisson bracket $\left\{ \cdot ,\cdot \right\} =%
\mathrm{i}\left[ \cdot ,\cdot \right] $ and the self-adjoint Vlasov
Hamiltonian $H(t,\rho )=H(t,\rho )^{\ast }$. On the manifold $M=\left\{ \psi
\psi ^{\ast }:\psi \in \mathcal{H}\right\} \subset \mathcal{L}$ of the
one-dimensional operators $\rho =\psi \psi ^{\ast }$ it takes the form 
\begin{equation}
\mathrm{i}\frac{\partial \psi (t)\psi (t)^{\ast }}{\partial t}=H(t,\psi
(t)\psi (t)^{\ast })\psi (t)\psi (t)^{\ast }-\psi (t)\psi (t)^{\ast
}H(t,\psi (t)\psi (t)^{\ast }),  \label{F}
\end{equation}%
which decomposes into a pair of mutually adjoint equations in $\mathcal{H}$
-- the abstract Hartree equations 
\begin{equation}
\mathrm{i}\frac{\partial \psi (t)}{\partial t}=H(t,\psi (t)\psi (t)^{\ast
})\psi (t),\text{\quad }-\mathrm{i}\frac{\partial \psi (t)^{\ast }}{\partial
t}=\text{ }\psi (t)^{\ast }H(t,\psi (t)\psi (t)^{\ast }).  \label{G}
\end{equation}%
This generalizes the Schr\"{o}dinger equation for the \emph{Hartree Hamilton
operators} $H(t,\psi \psi ^{\ast })$, the Vlasov Hamiltonians on $M$ that
depend on one-dimensional density $\psi \psi ^{\ast }$ in the space $%
\mathcal{H}$. Note that the equations (\ref{F}) describe a classical
field-Hamiltonian system on the space $\mathcal{H}$ with respect to the
symplectic form $2\func{Im}\left( \varphi |\psi \right) $, which determines
Poisson brackets on the submanifold $M\subset \mathcal{L}$: 
\begin{equation}
\left\{ \gamma ,\alpha \right\} _{cl}(\psi \psi ^{\ast })=2\func{Im}\left(
H(\psi \psi ^{\ast })\psi |A(\psi \psi ^{\ast })\psi \right) ,  \label{H}
\end{equation}%
where $A(\psi \psi ^{\ast })=\delta \alpha (\psi \psi ^{\ast })$.

\subsection{ \ }

A solution of equation (\ref{G}) can be written down implicitly for given
initial condition $\psi (t_{0})=\varphi $ by means of a time ordered mapping 
\cite{5}: 
\begin{equation}
\psi (t)=\exp \left\{ -\mathrm{i}\int_{t_{0}}^{t}\overset{s}{H}(s,\psi
(s)\psi (s)^{\ast })\mathrm{d}s\right\} \varphi \text{,}  \label{I}
\end{equation}%
which generalizes the ordinary Feynman time ordered product. Thus, the
solution of (\ref{C}) reduces in the given case to \textquotedblleft
disentangling\textquotedblright\ the expression (\ref{I}), i.e., to finding
an operator $V(t,t_{0},\varphi \varphi ^{\ast })$ by means of which (\ref{I}%
) can be represented explicitly as a function of $\varphi $: 
\begin{equation}
\psi (t)=V(t,t_{0},\varphi \varphi ^{\ast })\varphi  \label{J}
\end{equation}%
Sometimes, this can be readily done only for certain conditions $\varphi $.
For example, if for every $\nu =\varphi ^{\ast }\varphi $ there exists a
vector $\varphi =\varphi _{\nu }$ such that 
\begin{equation}
H(s,\varphi _{\nu }\varphi _{\nu }^{\ast })\varphi _{\nu }=\omega (s,\nu
)\varphi _{\nu }  \label{K}
\end{equation}%
directly for all $t$, then the solution (\ref{J}) has the form 
\begin{equation}
\psi (t)=\exp \left\{ -\mathrm{i}\int_{t_{0}}^{t}\omega (s,\nu )\mathrm{d}%
s\right\} \varphi _{\nu }\text{.}  \label{L}
\end{equation}%
This fact is a special case of the following assertion.

\begin{proposition}
\label{P2}Suppose Equation. (\ref{G}) has one or several integrals $\pi
_{j}(\psi \psi ^{\ast })$ of the form $\pi _{j}(\psi \psi ^{\ast })=\psi
^{\ast }P_{j}\psi $, where $P_{j}$ are self-adjoint commuting operators in $%
\mathcal{H}$, and suppose there exists an extremal $\psi =\varphi _{\mathbf{p%
}}$, $\mathbf{p}=\left\{ p_{j}\right\} $ of the functional $\gamma (t,\psi
\psi ^{\ast })$ which is the same for all $t$ and satisfies the conditions $%
\varphi _{\mathbf{p}}^{\ast }P_{j}\varphi _{\mathbf{p}}=p_{j}$, 
\begin{equation}
H(t,\varphi _{\mathbf{p}}\varphi _{\mathbf{p}}^{\ast })\varphi _{\mathbf{p}%
}=\sum_{j}\nu ^{j}P_{j}\varphi _{\mathbf{p}}\text{,}  \label{M}
\end{equation}%
where $\nu ^{j}$ are parameters conjugate to $p_{j}$ (Lagrangian
multipliers) determined by the conditions 
\begin{equation}
\nu ^{j}(t,\mathbf{p})=\frac{\partial h(t,\mathbf{p})}{\partial p_{j}}\text{%
,\quad }h(t,\mathbf{p})=\gamma (t,\varphi _{\mathbf{p}}\varphi _{\mathbf{p}%
}^{\ast })\text{.}  \label{N}
\end{equation}%
Then the solution of Equation. (\ref{G}) with the initial condition $\psi
(0)=\varphi _{\mathbf{p}}$ has the form 
\begin{equation}
\psi (t)=\exp \left\{ -\mathrm{i}\int_{t_{0}}^{t}\mathbf{v}(s,p)\mathbf{P}%
\mathrm{d}s\right\} \varphi _{\mathbf{p}}\text{,}  \label{O}
\end{equation}%
where $\mathbf{vP}=\sum \nu ^{j}P_{j}$.
\end{proposition}

This can be proved by direct substitution of (\ref{O}) into (\ref{C}) with
allowance for the invariance of $\gamma $ under the transformations $\psi
\mapsto \mathrm{e}^{\mathrm{i}\mathbf{qp}}\psi $ for all real parameters $%
\mathbf{q}=\left\{ q^{j}\right\} $. The parameters $q^{j}$, $\nu ^{j}$, $%
p_{j}$ have the meaning of generalized coordinates, velocities and momenta
respectively.

\begin{definition}
The solution (\ref{O}) of Equation. (\ref{G}) is called a generalized
soliton with momenta $p_{j}$.
\end{definition}

To the soliton (\ref{L}) there obviously corresponds the case of existence
of an extremal $\varphi _{\nu }$ for the integral $\pi _{0}(\psi \psi ^{\ast
})=\psi ^{\ast }\psi $ generated by the identity operator $P_{0}=I$.

Suppose, for example, the space $\mathcal{H}$ is the space of functions $%
\psi (\mathbf{x})$ on $\mathbb{R}^{d}\ni \mathbf{x}$ with scalar product $%
\left( \varphi |\psi \right) =\int \varphi \left( \mathbf{x}\right) 
\overline{\psi }\left( \mathbf{x}\right) \mathrm{d}\mathbf{x}$, and the
functional $\gamma $ is invariant with respect to the group of displacements:%
\begin{equation*}
\gamma (t,\psi _{\mathbf{q}}\psi _{\mathbf{q}}^{\ast })=\gamma (t,\psi \psi
^{\ast })\;\;\;\;\forall \psi _{\mathbf{q}}(\mathbf{x})=\psi (\mathbf{x}-%
\mathbf{q}),\mathbf{q}\in \mathbb{R}^{d}.
\end{equation*}%
In this case, the conditions of proposition (\ref{P2}) are satisfied by the
ordinary momenta%
\begin{equation*}
\pi _{j}(\psi \psi ^{\ast })=-\mathrm{i}\int \overline{\psi }(\mathbf{x}%
)\partial _{j}\psi (\mathbf{x})\mathrm{d}\mathbf{x},j=1,...,d,
\end{equation*}%
and one can expect that ordinary soliton solutions would exist - travelling
waves with momenta $\mathbf{p}=\left\{ p_{j}\right\} $:%
\begin{equation*}
\psi (\mathbf{x},t)=\varphi \left( \mathbf{x}-\int_{t_{0}}^{t}\mathbf{v}(s,%
\mathbf{p})\mathrm{d}s\right) .
\end{equation*}%
This last holds, for example, for the ordinary (scalar) Hartree-Fock
equation, for which the Hamilton operator $H(t,\psi \psi ^{\ast })$ has the
form 
\begin{equation}
H(\psi \psi ^{\ast })=-\partial ^{2}+\int \omega (\mathbf{x}-\mathbf{x}%
^{\prime })\left\vert \psi (\mathbf{x}^{\prime })\right\vert ^{2}\mathrm{d}%
\mathbf{x}^{\prime },  \label{P}
\end{equation}%
where $\partial ^{2}$ is the Laplacian with respect to $\mathbf{x}\in 
\mathbb{R}^{d}$, and where $\omega (\mathbf{x}-\mathbf{x}^{\prime })$ is the
potential of the two-body interaction, for example, the Green's function for
the equation $\partial ^{2}u(\mathbf{x})=4\pi \left\vert \psi (\mathbf{x}%
)\right\vert ^{2}$. Our treatment also encompasses more general cases of
Equation. (\ref{G}), for example, if $\partial ^{2}$ is a square with
respect to some indefinite metric in $\mathbb{R}^{d}$. An interesting case
arises when $\mathbb{R}^{d}$ is four-dimensional space-time, for which $%
\partial ^{2}$ is the d'Alembert operator $\square $.

\subsection{ \ }

In the general case, if there is a discrete basis in $\mathcal{H}$, it is
convenient to identify the vectors $\psi ^{\ast }$ and $\varphi $ with
covariant and contravariant tensors of the first rank $\psi _{x}$, $\varphi
^{x}=\overline{\varphi _{x}}$, fixing this basis. In this notation, the form 
$\psi ^{\ast }\varphi =\left( \varphi |\psi \right) $ is simply the tensor
contraction $\psi _{x}\varphi ^{x}$ such that the mapping $\psi \mapsto \psi
^{\ast }$ is the operation of complex conjugation and lowering the
superscript by means of an antilinear metric tensor $J:\psi _{x}=J_{xy}\psi
^{y},\psi ^{x}=J^{xy}\psi _{y}$. At the same time, the operators $\rho \in 
\mathcal{L}$, $A\in \mathcal{A}$ are represented by matrices $\rho
_{y}^{\,\;x}$, $A_{x}^{\;y~}$, the bilinear form $\left\langle \rho
,A\right\rangle $ is the contraction $\rho _{y}^{\,\;x}A_{x}^{\;y}$, and the
derivative $H(\rho )=\delta \gamma (\rho )$ is given by the partial
derivatives $H_{x}^{\;y}(\rho )=\frac{\partial \gamma (\rho )}{\partial \rho
_{y}^{\,\;x}}$. Further, we shall encounter tensor powers $\rho ^{\otimes n}$%
, $\delta ^{\otimes n}$, which in our notation are the products $\rho
_{y_{1}}^{\,\;x_{1}}...\rho _{y_{n}}^{\,\;x_{n}}$, $\delta
_{x_{1}}^{\;y_{1}}...\delta _{x_{n}}^{\;y_{n}}=\frac{\partial ^{n}}{\partial
\rho _{y_{1}}^{\,\;x_{1}}...\partial \rho _{x_{n}}}$. Formally, the tensor
notation can also be used even in the continuous index case if the partial
derivatives are understood as variational derivatives and contraction as an
integral with respect to a given measure; however the matrix elements are
then represented by generalized functions.

\section{Uniformization and Second Quantization}

\subsection{ \ }

We shall say that a Hamiltonian system (\ref{C}) is uniformizable if the
Hamilton functional $\gamma (t,\rho )$ is a polynomial: 
\begin{equation}
\gamma (t,\rho )=\sum_{n=1}^{N}\frac{1}{n!}\left\langle \rho ^{\otimes
n},W^{(n)}(t)\right\rangle  \label{Q}
\end{equation}%
or entire function $(N=\infty )$. Here, $\rho ^{\otimes n}$ are tensor
powers of the element $\rho $ $\in \mathcal{L}$ generating the symmetrical
tensor-spaces $\mathcal{L}_{n}$\footnote{%
That is $\mathcal{L}_{n}$ is the minimal Fr\'{e}shet subspace of the
projective tensor power $\mathcal{L}^{\otimes n}$ containing all $\rho
^{\otimes n}$, $\rho \in \mathcal{L}$.}, $W^{(n)}(t)$ are elements of the
dual algebras $\mathcal{A}^{(n)}=\mathcal{L}_{n}^{\ast }$ , and $%
\left\langle \rho ^{\otimes n},A^{(n)}\right\rangle $ are linear forms on $%
\mathcal{A}^{(n)}\ni A^{(n)}$ equal to the products $\left\langle \rho
,A_{1}\right\rangle ...\left\langle \rho ,A_{n}\right\rangle $ on the $%
\mathcal{A}^{(n)}$-generating elements $A^{(n)}=A_{1}\otimes ...\otimes
A_{1}\equiv A_{1}^{\otimes n}$. Note that $W^{(n)}$ are represented as the
symmetric tensors in any discrete basis, and are uniquely determined from $%
\gamma (\rho )$ as $n$-th derivatives $\delta ^{\left( n\right) }\gamma
(\rho )$ at the point $\rho =0$ defined as the tensor power $\delta
^{(n)}\equiv \delta ^{\otimes n}$ of the derivative $\delta $ in (\ref{A})
such that%
\begin{equation*}
\delta ^{\otimes n}\mathrm{e}^{\left\langle \rho ,A\right\rangle }=\mathrm{e}%
^{\left\langle \rho ,A\right\rangle }A^{\otimes n}
\end{equation*}%
for any $A\in \mathcal{A}$. The symmetrical spaces $\mathcal{A}^{(n)}$ like
the full tensor power spaces $\mathcal{A}^{\otimes n}$ are Hamiltonian
(Lie-Jordan) algebras with respect to products $H^{(n)}\cdot A^{(n)}$, $%
\left\{ H^{(n)},A^{(n)}\right\} $ defined on the primitive elements $%
H^{(n)}=\bigotimes_{i=1}^{n}H_{i}$, $A^{(n)}=\bigotimes_{i=1}^{n}A_{i}$ as 
\begin{equation}
H^{(n)}\cdot A^{(n)}=\bigotimes_{i=1}^{n}H_{i}\cdot A_{i},\quad \left\{
H^{(n)},A^{(n)}\right\} =\sum_{j=1}^{n}\bigotimes_{\substack{ i=1  \\ i\neq
j }}^{n}H_{i}\cdot A_{i}\otimes \left\{ H_{j},A_{j}\right\}  \label{U}
\end{equation}%
such that $H_{1}^{\otimes n}\cdot A_{1}^{\otimes n}=\left( H_{1}A_{1}\right)
^{\otimes n}$, $\left\{ H_{1}^{\otimes n},A_{1}^{\otimes n}\right\}
=nH_{1}^{\otimes \left( n-1\right) }\cdot A_{1}^{\otimes \left( n-1\right)
}\otimes _{s}\left\{ H_{1},A_{1}\right\} $.

The symmetry of $W^{(n)}(t)$ makes it possible to write the Vlasov
Hamiltonian $H(\rho )=$ $\delta \gamma (\rho )$ determining the Vlasov
equation (\ref{C}), in the form 
\begin{equation}
H(t,\rho )=\sum_{n=1}^{N-1}\frac{1}{n!}\left\langle \rho ^{\otimes
n},W^{(n+1)}(t)\right\rangle \text{,}  \label{R}
\end{equation}%
where $\left\langle \rho ^{\otimes n},A^{(n+1)}\right\rangle $, the elements
of the algebra $\mathcal{A}$, are equal to $\left\langle \rho
,A_{1}\right\rangle ^{n}A$ for $A^{(n+1)}=A_{1}\otimes ...\otimes A_{1}$. An
example of such an expression with $N=2$ gives the operator (\ref{P}) if $%
\psi \psi ^{\ast }$ is replaced by an arbitrary density operator $\rho =\sum
\lambda _{i}\psi _{i}\psi _{i}^{\ast }$.

\begin{definition}
Let $\varepsilon >0$ be a parameter; an $\varepsilon $-uniformization of the
classical Hamiltonian system (\ref{R}) with the Hamiltonian functional $%
\gamma \left( t,\rho \right) $ is the system of the Hamiltonians 
\begin{equation}
H^{(n)}(t)=\frac{1}{\varepsilon }(1+\delta \varepsilon )^{\otimes n}\gamma
(t,\rho )\!{}\mid _{\rho =0}=\sum_{m=1}\binom{n}{m}\varepsilon ^{m-1}\left(
I^{(n-m)}\otimes _{s}W^{(m)}(t)\right)  \label{S}
\end{equation}%
as the real elements in the Hamiltonian algebras $\mathcal{A}^{(n)}$. Here $%
I^{(n-m)}\otimes _{s}W^{(m)}$ is the symmetrization of the tensor products $%
I^{(n-m)}\otimes W^{(m)}$ of the identities $I^{(n-m)}=I\otimes ...\otimes I$
of the algebras $\mathcal{A}^{(n-m)}$ and the elements $W^{(m)}\in $ $%
\mathcal{A}^{(m)}$.
\end{definition}

For example, the uniformization of the Hamilton operators (\ref{P}) of the
Hartree equation (\ref{G}) leads to the system of operators 
\begin{equation}
H^{(n)}=-\sum_{i=1}^{n}\partial _{i}^{2}+\varepsilon
\sum_{i_{1}=1}^{n}\sum_{i_{2}=1}^{i_{1}-1}\omega (\mathbf{x}_{i_{1}}-\mathbf{%
x}_{i_{2}})\text{,}  \label{T}
\end{equation}%
where $\partial _{i}^{2}$ are the Laplace operators with respect to the
variables $\mathbf{x}_{i}\in \mathbb{R}_{i}^{d}$. For four-dimensional
space-time $\mathbb{R}^{4}$, for which $\partial ^{2}$ is the d'Alembert
operator, the uniformized Hamiltonians are ultrahyperbolic operators for $%
n\geqslant 2$.

\subsection{ \ }

Note that the Hamiltonians (\ref{S}) determine in the spaces $\mathcal{L}%
_{n} $ the linear equations 
\begin{equation}
\frac{\partial \rho _{n}(t)}{\partial t}=\left\{ \rho
_{n}(t),H^{(n)}(t)\right\}  \label{V}
\end{equation}%
which are abstract Liouville (or von Neumann) equations understood as
predual to the abstract Heisenberg equations 
\begin{equation}
\frac{\partial A^{(n)}(t)}{\partial t}=\left\{ H^{(n)}(t),A^{(n)}(t)\right\}
,  \label{W}
\end{equation}%
so that the relation $\left\langle \rho _{n}(t),A_{0}^{(n)}\right\rangle
=\left\langle \rho _{n}^{0},A^{(n)}(t)\right\rangle $ holds for all initial $%
A_{0}^{(n)}=A^{(n)}(t_{0})\in \mathcal{A}^{(n)}$ and $\rho _{n}^{0}=\rho
_{n}(t_{0})\in \mathcal{L}_{n}$.

In order to develop a compact functional derivative representation of all
these independent equations and find their relation to the abstract Vlasov
equation we introduce the representing functionals 
\begin{equation}
\alpha (\rho )=\exp \left\{ -\frac{1}{\varepsilon }\left\langle \rho
,I\right\rangle \right\} \sum \frac{1}{n!\varepsilon ^{n-1}}\left\langle
\rho ^{\otimes n},A^{(n)}\right\rangle \text{,}  \label{X}
\end{equation}%
by means of which the elements $A^{(n)}\in \mathcal{A}^{(n)}$ can be
calculated as the derivatives 
\begin{equation}
A^{(n)}=\varepsilon ^{n-1}\delta ^{\otimes n}\exp \left\{ \frac{1}{%
\varepsilon }\left\langle \rho ,I\right\rangle \right\} \alpha (\rho )\mid
_{\rho =0}=\frac{1}{\varepsilon }(1+\varepsilon \delta )^{\otimes n}\alpha
(\rho )\!{}\mid _{\rho =0}\text{.}  \label{Y}
\end{equation}%
Here, $A^{(0)}$ is a complex number $A^{(0)}\in \mathbb{C}\equiv \mathcal{A}%
^{(0)}$.

\begin{proposition}
The system of equations (\ref{W}) can be expressed in the form of the
pseudodifferential equation 
\begin{equation}
\frac{\partial \alpha (t,\rho )}{\partial t}=\left\langle \left\{ \overset{2}%
{\rho },H\left( t,\overset{2}{\rho }\cdot \left( I+\varepsilon \overset{1}{%
\delta }\right) \right) \right\} ,\overset{1}{\delta }\right\rangle \alpha
(t,\rho ),  \label{Z}
\end{equation}

defined in terms of the symbol-Hamiltonian 
\begin{equation*}
H\left( t,\rho \cdot \left( I+\varepsilon A\right) \right) \equiv \exp
\left\{ \varepsilon \left\langle \overset{2}{\rho },\overset{1}{\delta }%
\cdot A\right\rangle \right\} H(t,\rho ),
\end{equation*}
where the superscripts indicate the order of application of the operators of
(tensor) multiplication by $\rho $ and the derivation with respect to $\rho $%
. The equation (\ref{Z}) has the limit at $\varepsilon \longrightarrow 0$ of
first order differential equation%
\begin{equation*}
\frac{\partial }{\partial t}\alpha (t,\rho )=\left\langle \left\{ \rho
,H\left( t,\rho \right) \right\} ,\delta \right\rangle \alpha (t,\rho )
\end{equation*}
giving the weak form of the Vlasov equation for $\alpha \left( t,\rho
\right) =\left\langle \rho \left( t\right) ,A\right\rangle $.
\end{proposition}

The proof of this proposition is contained in the derivation of the formula
(expansions) (\ref{AC}) and (\ref{AD}) in the Appendix. Here we only note
that the symbol 
\begin{equation*}
G(t,\rho ,A)=\left\langle \left\{ \rho ,H\left( t,\rho \cdot \left(
I+\varepsilon A\right) \right) \right\} ,A\right\rangle
\end{equation*}%
of the equation (\ref{Z}), which is determined \ on the exponential
functionals $\alpha \left( \rho \right) =\exp \left\langle \rho
,A\right\rangle $ by 
\begin{equation}
G\left( t,\overset{2}{\rho },\overset{1}{\delta }\right) \mathrm{e}%
^{\left\langle \rho ,A\right\rangle }=\mathrm{e}^{\left\langle \rho
,A\right\rangle }G\left( t,\rho ,A\right) \text{,}  \label{AA}
\end{equation}%
depends analytically on $\varepsilon $ as a power series expansion in $%
\varepsilon $. Comparing (\ref{Z}) with (\ref{D}), we readily see that for $%
\varepsilon =0$ the derivative (\ref{Z}) simply coincides with the
derivative of $\alpha (t,\rho )=\alpha (\rho (t))$ along the trajectories of
the Vlasov equation (\ref{C}). Expanding the operator $G\left( t,\rho
,\delta \right) $ in powers of $\varepsilon $ and substituting in Equation. (%
\ref{Z}) the solution in the form of the series $\alpha (t,\rho
)=\sum_{k=0}^{\infty }\varepsilon ^{k}\alpha _{k}(t,\rho )$, we obtain by
the method of successive approximation the following system of coupled
equations: 
\begin{equation}
\frac{\partial \alpha _{k}(t)}{\partial t}=\sum_{n=0}^{k}\frac{1}{\left(
n+1\right) !}\left\langle \left\{ \left( \overset{2}{\rho }\right) ^{\otimes
\left( n+1\right) },H^{\left( n+1\right) }(t,\rho )\right\} ,\left( \overset{%
1}{\delta }\right) ^{\otimes \left( n+1\right) }\right\rangle \alpha
_{k-n}(t)\text{,}  \label{AB}
\end{equation}%
where $H^{(m)}(t,\rho )=\delta ^{\otimes m}\gamma (t,\rho )$. Thus, the
Vlasov equation (\ref{C}) is a Hamiltonian system of the characteristics for
the zeroth-order approximation $\alpha _{0}(t)$ with initial condition $%
\alpha (t_{0})=\alpha _{0}$.

\subsection{ \ }

In general, the uniformization procedure can be regarded as a linear functor
from the classical algebra of entire Hamilton functions $\gamma (\rho )$
with ordinary product $\gamma (\rho )\alpha (\rho )$ and Poisson bracket (%
\ref{E}) to a new Lie-Jordan (Hamiltonian) algebra $\mathfrak{A}$ of these
functions with not necessarily associative product 
\begin{eqnarray}
\left( \gamma \cdot \alpha \right) \left( \rho \right) &=&\frac{1}{%
\varepsilon }\gamma \left( \rho \cdot \left( I+\varepsilon \overset{1}{%
\delta }\right) \right) \alpha \left( \rho \right)  \label{AC} \\
&=&\sum_{n=0}^{\infty }\frac{\varepsilon ^{n-1}}{n!}\left\langle \rho
^{\otimes n},\delta ^{\otimes n}\gamma \left( \rho \right) \cdot \delta
^{\otimes n}\alpha \left( \rho \right) \right\rangle  \notag
\end{eqnarray}%
and uniformized Poisson brackets 
\begin{eqnarray}
\left\{ \gamma ,\alpha \right\} \left( \rho \right) &=&\left\langle \rho
,\left\{ \delta \gamma \left( \rho \cdot \left( I+\varepsilon \delta \right)
\right) ,\delta \alpha \left( \rho \right) \right\} \right\rangle  \label{AD}
\\
&=&\sum_{n=0}^{\infty }\frac{\varepsilon ^{n-1}}{n!}\left\langle \rho
^{\otimes n},\left\{ \delta ^{\otimes n}\gamma \left( \rho \right) ,\delta
^{\otimes n}\alpha \left( \rho \right) \right\} \right\rangle \text{.} 
\notag
\end{eqnarray}%
The latter in accordance with (\ref{B}) determines for a fixed Hamilton
functional $\gamma (t,\rho )$ the \emph{uniformized Hamiltonian derivation} (%
\ref{Z}) on the algebra $\mathfrak{A}$, in which $H(t,\rho )=$ $\delta
\gamma (t,\rho )$, since Equations (\ref{X}) and (\ref{Y}) establish an
isomorphism between the algebra $\mathfrak{A}$ and the direct sum algebra $%
\bigoplus_{n=0}^{\infty }\mathcal{A}^{(n)}$ of symmetrized tensor powers $%
\mathcal{A}^{(n)}$ of the algebra of elementary observables $\mathcal{A}$
since the products (\ref{AC}) and (\ref{AD}) are simply representing
functionals for the products $H^{(n)}\cdot A^{(n)}$, $\left\{
H^{(n)},A^{(n)}\right\} $ of components of the elements $\hat{A}=$ $%
\bigoplus_{n=0}^{\infty }A^{(n)}$ and $\hat{H}=$ $\bigoplus_{n=0}^{\infty
}H^{(n)}$ (this last follows from the proof of Proposition (\ref{P1})). In
the limit $\varepsilon \mapsto 0$, the Jordan product (\ref{AC}) multiplied
by $\varepsilon $ goes over into the ordinary pointwise product $\gamma
(\rho )\alpha (\rho )$, and the Lie product (\ref{AD}) in this limit
obviously coincides with the classical Poisson brackets (\ref{E}).

In the case when the Jordan product in $\mathcal{A}$ is associative, the
uniformized product (\ref{AC}) also remains associative. This is the case
when $\mathcal{A}$ is the classical algebra of single particle observables $%
A(\mathbf{q},\mathbf{p})$. Then the uniformization is equivalent to the
transition from the classical field Hamiltonian system described by the
Vlasov equation itself to the classical Hamiltonians (\ref{U}), where $%
\varepsilon $ is the coupling constant. In this sense, the uniformization
procedure is the inverse of the procedure of derivation of the Vlasov
equation from the Bogolyubov equation \cite{6}. One could naturally call
this replacement of the classical field system by a system with variable
particle number as quantization, however, we shall avoid this terminology
here since by quantization one always understands transitions from a
classical to a definitely nonclassical system, whereas the example
considered here shows that the uniformized system may remain classical.

The same interpretation remains valid in the quantum case when $\mathcal{A}$
is an algebra of operators in Hilbert single-particle space $\mathcal{H}$.
The uniformized system of a variable number of identical quantized particles
is, in contrast to the original field classical system, no longer classical,
and in this sense the described procedure can be regarded as a certain
quantization, and it can be given a Heisenberg from with generating
commutation relations 
\begin{equation}
\left[ \left\langle \rho ,H\right\rangle ,\left\langle \rho ,A\right\rangle %
\right] =\left\langle \rho ,\left[ H,A\right] \right\rangle \text{,}
\label{AE}
\end{equation}%
if one introduces an associative and noncommutative product $\left( \gamma
\cdot \alpha \right) \left( \rho \right) $ as the representing functional
for the products $H^{(n)}\cdot A^{(n)}$ of operators in the tensor powers $%
\mathcal{H}^{(n)}$ of the Hilbert space $\mathcal{H}$. At the same time, the
complex extension of $\mathcal{A}$ is a linear noncommutative associative
algebra isomorphic to the algebra of operators $\hat{A}\in \mathfrak{H}=$ $%
\bigoplus_{n=0}^{\infty }\mathcal{H}^{(n)}$ of decomposed form $\hat{A}%
=\oplus A^{(n)}$, where $A^{(n)}$ are symmetric tensor operators in $%
\mathcal{H}^{(n)}$ that leave invariant the subspaces $\mathcal{H}_{\pm
}^{(n)}\subset \mathcal{H}^{(n)}$ of the completely symmetric $\left( 
\mathcal{H}_{+}^{(n)}\right) $, as well as the completely antisymmetric $%
\left( \mathcal{H}_{-}^{(n)}\right) $ tensors $\psi ^{(n)}\in \mathcal{H}%
^{(n)}$. This last means that the elements of the algebra $\mathcal{A}$ can
be interpreted as the observables of a second quantized system of identical
bosons or fermions depending on whether or not they are regarded as
operators $\hat{A}$ in Fock spaces $\mathfrak{H}_{\pm }=$ $%
\bigoplus_{n=0}^{\infty }\mathcal{H}_{\pm }^{(n)}$ that are symmetric or
antisymmetric under permutations of particles. Such Fock representations of
the algebra $\mathcal{A}$ are obtained by means of the functionals (\ref{W})
in accordance with the formula $A=\frac{1}{\varepsilon }\alpha \left(
\varepsilon \overset{1}{a_{\pm }}\overset{2}{a_{\pm }^{\ast }}\right) $,
where $\hat{a}_{\pm }$, $\hat{a}_{\pm }^{\ast }$ are vector operators of
annihilation and creation of bosons (with plus sign) and fermions (with
minus sign) acting on the Fock spaces $\mathfrak{H}_{+}$, $\mathfrak{H}_{-}$
respectively, and the numbers $1$ and $2$ above the operators indicate the
normal order, in accordance with which the monomials $\left\langle \left( 
\overset{1}{a}\overset{2}{a^{\ast }}\right) ^{\otimes
n},A^{(n)}\right\rangle $ are written as the normal products $\hat{a}^{\ast
\otimes n}A^{(n)}\hat{a}^{\otimes n}$. The linear observables $\alpha \left(
\rho \right) =\left\langle \rho ,A\right\rangle $ as functions on $\mathcal{A%
}\ni A$ are then represented in $\mathfrak{H}_{\pm }$ by operator-valued
forms $\hat{n}(A)=\hat{a}_{\pm }^{\ast }A\hat{a}_{\pm }$, and the
commutation relations (\ref{AE}) can be deduced directly from the canonical
commutation relations for the operators $\hat{a}_{\pm }$, $\hat{a}_{\pm
}^{\ast }$: 
\begin{equation}
\left[ \hat{n}(H),\hat{n}(A)\right] =\left[ \hat{a}_{\pm }^{\ast }H\hat{a}%
_{\pm },\hat{a}_{\pm }^{\ast }A\hat{a}_{\pm }\right] =\hat{a}_{\pm }^{\ast }%
\left[ H,A\right] \hat{a}_{\pm }=\hat{n}\left( \left[ H,A\right] \right) 
\text{.}  \label{AF}
\end{equation}%
Hamiltonians $\hat{H}=\frac{1}{\varepsilon }\gamma \left( \varepsilon 
\overset{1}{a}\overset{2}{a^{\ast }}\right) $ of the nonlinear form (\ref{Q}%
) can be represented in the Fock spaces $\mathfrak{H}_{\pm }$ by expansions
in the coupling constant $\varepsilon $: 
\begin{equation}
\hat{H}(t)=\sum_{n=1}^{N}\frac{\varepsilon ^{n-1}}{n!}\hat{a}_{\pm }^{\ast
\otimes n}W^{(n)}(t)\hat{a}_{\pm }^{\otimes n}\text{.}  \label{AG}
\end{equation}%
Thus, uniformization of the quantum Vlasov equation can be regarded as the
algebraic basis for a simplified method of second quantization which avoids
explicit introduction of unphysical - odd - observables (which do not
commute with the operator of the total particle number $n\left( \mathbf{\hat{%
I}}\right) $) and takes into account in a unified manner for Bose and Fermi
systems the principle of indistinguishability of identical particles. The
calculus (\ref{AC}-\ref{AD}) of the representing functionals (\ref{X})
together with Equation. (\ref{Y}) combines the holomorphic calculus of the
operators $\hat{A}=\bigoplus A^{(n)}$ in the fock spaces $\mathfrak{H}_{\pm
} $ developed by Bargmann \cite{10} for the case of Bose-Einstein statistics
and Berezin \cite{11} for Fermi-Dirac statistics; at the same time, the
commutation relations (\ref{AE}) replace the canonical commutation relations
for the operators $\hat{a}_{\pm }$, $\hat{a}_{\pm }^{\ast }$.

All that we have said remains true in the more general case when the metric
of the space $\mathcal{H}$ can be indefinite except for the interpretation
of uniformization as second quantization since the spaces $\mathfrak{H}%
=\bigoplus \mathcal{H}^{(n)}$, like their subspaces $\mathfrak{H}_{\pm }$
need not be Hilbert spaces, but only, like $\mathcal{H}$, pseudo-Hilbert
spaces. The representations $\bigoplus \mathcal{A}^{(n)}$ of the algebra $%
\mathcal{A}$ by the operators (\ref{X}) in the spaces $\mathcal{H}^{(n)}$
are then representations with indefinite metric, and are not Fock
representations. For this reason, the dual elements $\rho
=\bigotimes_{n}\rho _{n}$, where $\rho _{n}\in \mathcal{L}_{n}$ are
operators of the form $\rho _{n}=\sum_{i}\lambda _{i}\psi _{i}^{(n)}\psi
_{i}^{(n)\ast }$, cannot, in general, be interpreted as statistical
operators of $n$-particle states even if $\lambda _{i}\geqslant 0$ and $%
\sum_{i}\lambda _{i}=1$. Nevertheless, investigation of the linear equations
(\ref{S}) or the (abstract) Schr\"{o}dinger equations 
\begin{equation}
\mathrm{i}\frac{\partial \psi ^{(n)}(t)}{\partial t}=H^{(n)}\psi
^{(n)}(t),\quad -\mathrm{i}\frac{\partial \psi ^{(n)}(t)^{\ast }}{\partial t}%
=\psi ^{(n)}(t)^{\ast }H^{(n)}\text{,}  \label{AH}
\end{equation}%
into which they decompose for initial conditions of the one-dimensional form 
$\rho _{n}(t_{0})=\psi ^{(n)}\psi ^{(n)}{}^{\ast }$, enables one to extract
the entire information about the dynamics of the uniformized system. It is
sufficient to consider only the symmetric or antisymmetric $n$-tensors $\psi
^{(n)}\in \mathcal{H}^{(n)}$, for which the products $\psi ^{(n)}\psi
^{(n)}{}^{\ast }$ are elements of the space $\mathcal{L}_{n}$ of symmetric $%
n $-tensors $\rho _{n}$ since equations (\ref{AC}) preserve the symmetry of $%
\psi ^{(n)}$ by virtue of the fact that the Hamiltonians $H^{(n)}\in 
\mathcal{A}^{(n)}$ commute with the permutation operators.

\section{Uniformized Solutions of the Hartree Equation}

\subsection{ \ }

Here, we consider a method based on uniformization for disentangling the
Time ordered mapping (\ref{I}), this giving solutions of the Hartree
equation (\ref{G}) with Hamiltonian of the form (\ref{R}), where $\rho =\psi
\psi ^{\ast }$ and $\left\langle \rho ^{\otimes n},W^{(n+1)}\right\rangle =$ 
$\psi ^{\ast \otimes n}W^{(n+1)}\psi ^{\otimes n}$. To be specific, we shall
assume throughout that the forms $\psi ^{\ast \otimes n}A^{(n)}\psi
^{\otimes n}$\ are c-numbers, although they could also be regarded as even
elements of a Grassmann algebra with $\psi $ and $\psi ^{\ast }$ regarded as
anticommuting generators. At the same time, $\psi ^{\ast \otimes
n}A^{(n)}\psi ^{\otimes n}=\psi ^{\ast }A_{1}\psi \cdots \psi ^{\ast
}A_{n}\psi $ for $A^{(n)}=\bigoplus_{i=1}^{\infty }A_{i}$, and $\psi ^{\ast
}A\psi =\left( A\varphi |\psi \right) $.

We shall regard the differences $H^{(n+1)}(t)-H^{(n)}(t)\otimes \mathbf{I}$
as operators in $\mathcal{H}_{+}^{(n)}\otimes \mathcal{H}$ and denoted by $%
\Omega ^{(n+1)}(t,t_{0})$ the operators in $\mathcal{H}_{+}^{(n)}\otimes 
\mathcal{H}$ obtained by there transformation: 
\begin{equation}
\Omega ^{(n+1)}(t,t_{0})=U^{(n)}(t,t_{0})^{\ast }\left(
H^{(n+1)}(t)-H^{(n)}(t)\otimes \mathbf{I}\right) U^{(n)}(t,t_{0})\text{,}
\label{AI}
\end{equation}%
where $U^{(n)}(t,t_{0})$ are unitary operators in $\mathcal{H}_{+}^{(n)}$
determining the solutions of equations (\ref{S}): 
\begin{equation}
U^{(n)}(t,t_{0})=\exp \left\{ -\mathrm{i}\int_{t_{0}}^{t}\overset{s}{H}%
^{(n)}(s)\mathrm{d}s\right\} \text{.}  \label{AJ}
\end{equation}

\begin{definition}
We give the name $\varepsilon $ solution of equation (\ref{G}) with the
operator (\ref{R}) and initial condition $\psi (t_{0})=\psi $ to the
following expression $\psi (t)=V(t,t_{0},\varphi \varphi ^{\ast })\varphi $,
where 
\begin{equation}
V(t,t_{0},\varphi \varphi ^{\ast })\varphi =\exp \left\{ -\frac{1}{%
\varepsilon }\varphi ^{\ast }\varphi \right\} \sum_{n=0}^{\infty }\frac{%
\varepsilon ^{n-1}}{n!}\varphi ^{\ast \otimes n}V^{(n+1)}(t,t_{0})\varphi
^{\otimes n}\text{,}  \label{AK}
\end{equation}%
and $V^{(n+1)}(t,t_{0})$ are unitary operators in $\mathcal{H}%
_{+}^{(n)}\otimes \mathcal{H}$ with generators \ (\ref{AI}), i.e., the
time-ordered products 
\begin{equation}
V^{(n+1)}(t,t_{0})=\exp \left\{ -\mathrm{i}\int_{t_{0}}^{t}\overset{s}{%
\Omega }^{(n+1)}(s,t_{0})\mathrm{d}s\right\} \text{.}  \label{AL}
\end{equation}
\end{definition}

The justification for this definition is the following proposition, which
reduces the problem of disentangling the Time-ordered mapping (\ref{I}) for
uniformized systems to calculation of Time-ordered products (\ref{AL}).

\begin{proposition}
The solution of the uniformized equation (\ref{G}) with any initial
condition $\psi (t_{0})=\varphi $ coincides with the $\varepsilon
\longrightarrow 0$ limit of the corresponding $\varepsilon $ solution.
\end{proposition}

This proposition is readily proved by noting that the Time-ordered products (%
\ref{AL}) are found by disentangling the operator $\left( U^{(n)}\otimes 
\mathbf{I}\right) ^{-1}$ with the generator $H^{(n)}\otimes \mathbf{I}$ from
the operator $U^{(n+1)}$ with generator $H^{(n+1)}=H^{(n)}\otimes \mathbf{I+}%
\Omega ^{(n+1)}$, 
\begin{equation}
V^{(n+1)}(t,t_{0})=\left( U^{(n)}\left( t,t_{0}\right) \otimes \mathbf{I}%
\right) ^{-1}U^{(n+1)}\left( t,t_{0}\right) \text{.}  \label{AM}
\end{equation}%
This, the $\varepsilon $ solution essentially reduces to finding the unitary
operators (\ref{AJ}) representing the solutions of the Schr\"{o}dinger
equation (\ref{AH}) in the form $\psi ^{(n)}(t)=U^{(n)}(t,t_{0})\varphi
^{(n)}$ for any initial condition $\varphi ^{(n)}\in \mathcal{H}_{+}^{(n)}$.
The operators (\ref{AM}) multiplied by $\sqrt{n}$ are none other than the
nonzero matrix elements of the vector operators of annihilation of bosons $%
\hat{a}(t)$ in the Heisenberg representation, and the $\varepsilon $
solution as a function of $\varphi $, $\varphi ^{\ast }$ is the symbol of
the vector operator $\sqrt{\varepsilon }\hat{a}(t)=\hat{\psi}(t)$ ordered
normally with respect to the initial operators $\hat{\varphi}=\sqrt{%
\varepsilon }\hat{a}_{+}(t_{0})$. For $\varepsilon =0$, the initial vector
operators $\hat{\varphi}$ coincide with the numerical vectors $\varphi $,
and the $\varepsilon $ solutions coincide with the solutions of the Hartree
equation since the Heisenberg equation for the symbol $\hat{\psi}(t)$ in the
representation of the instantaneous time $t$ coincides exactly with the
Hartree equation.

\subsection{ \ }

The disentangling of the Time-ordered mappings in this way requires the
solution of all the Schr\"{o}dinger equations (\ref{AH}) for $n=1,2,...$. It
is of interest to ask whether one can find not all but at least some
solutions of the Hartree equation by solving not all but, for example, two
neighboring equations (\ref{AH}).

Let $\varphi ^{\left( n\right) }$, $\varphi ^{\left( n+1\right) }$ be
(generalized) stationary solutions of Equations. (\ref{AH}) corresponding to
the eigenvalues $\lambda _{n}(t)$, $\lambda _{n+1}(t)$ of the uniformized
Hamiltonians $H^{(n)}(t)$, $H^{(n+1)}(t)$. One can evidently write the
matrix element $\psi (t)=\varphi ^{\left( n\right) \ast
}V^{(n+1)}(t,t_{0})\varphi ^{\left( n+1\right) }$ of the operator (\ref{AM})
in the form (\ref{L}), where 
\begin{equation}
\varphi ^{\left( n\right) \ast }\varphi ^{\left( n+1\right) }=\varphi
_{v},\quad v=\varepsilon n,\quad \omega (t,v)=\lambda _{n+1}(t)-\lambda
_{n}(t)\text{.}  \label{AN}
\end{equation}%
If all quantities (\ref{AN}) have a limit as $\varepsilon \longrightarrow 0$%
, then one can expect that in this limit the given matrix element defines a
certain stationary solution of the Hartree equation, while for $\varepsilon
\neq 0$ it gives an approximate solution of this equation.

Generally speaking, suppose the Hartree equation (\ref{G}) has one or
several integrals of quadratic form $\varphi ^{\ast }P_{j}\varphi $,
including $\mathbf{P}_{0}=\mathbf{I}$, and suppose $\left\vert \mathbf{p}%
\right\rangle $ are (generalized) eigenvectors of the operators $\varepsilon 
\hat{a}^{\ast }P_{j}\hat{a}$ and the Hamilton operator $\gamma \left(
\varepsilon \overset{1}{a}\overset{2}{a}^{\ast }\right) $ in the space $%
\mathfrak{H}=\bigoplus \mathcal{H}_{+}^{(n)}$ corresponding to the
eigenvalues $p_{j}$ and $h(t,\mathbf{p})$. We denote the integrals of the
matrix elements of the Heisenberg vector operators $\hat{a}(t)$ by 
\begin{equation}
\psi _{\mathbf{p}}(t)=\sqrt{\varepsilon }\int \left\langle \mathbf{p}%
\right\vert \hat{a}(t)\left\vert \mathbf{p}^{\prime }\right\rangle \mathrm{d}%
\mathbf{p}^{\prime }\text{.}  \label{AO}
\end{equation}

\begin{proposition}
The vectors (\ref{AO}) can be expressed in terms of the initial vectors $%
\varphi _{\mathbf{p}}=\psi (t_{0})$ in accordance with 
\begin{equation}
\psi _{\mathbf{p}}(t)=\exp \left\{ -\frac{\mathrm{i}}{\varepsilon }%
\int_{t_{0}}^{t}\left( h\left( s,\mathbf{p}+\varepsilon \mathbf{P}\right)
-h\left( s,\mathbf{p}\right) \right) \mathrm{d}s\right\} \varphi _{\mathbf{p}%
}\text{.}  \label{AP}
\end{equation}

\begin{proof}
For any Heisenberg operator $\hat{A}(t)$ we introduce in the space $%
\mathfrak{H}$ the function 
\begin{equation}
\alpha \left( t,\mathbf{p},\mathbf{q}\right) =\int \exp \left\{ \frac{%
\mathrm{i}}{\varepsilon }\mathbf{q\cdot }\left( \mathbf{p-p}^{\prime
}\right) \right\} \left\langle \mathbf{p}\right\vert \hat{A}(t)\left\vert 
\mathbf{p}^{\prime }\right\rangle \mathrm{d}\mathbf{p}^{\prime }\text{.}
\label{AQ}
\end{equation}%
It is easy to see that the Heisenberg equations (\ref{W}) lead to the
following equation for the function (\ref{AQ}): 
\begin{equation}
\mathrm{i}\frac{\partial \alpha \left( t,\mathbf{p},\mathbf{q}\right) }{%
\partial t}=\frac{1}{\varepsilon }\left( h\left( t,\mathbf{p}+\mathrm{i}%
\varepsilon \frac{\partial }{\partial \mathbf{q}}\right) +h\left( t,\mathbf{p%
}\right) \right) \alpha \left( t,\mathbf{p},\mathbf{q}\right) \text{.}
\label{AR}
\end{equation}%
In particular, for the vector function 
\begin{equation}
\psi _{\mathbf{p,q}}(t)=\int \exp \left\{ \frac{\mathrm{i}}{\varepsilon }%
\mathbf{q\cdot }\left( \mathbf{p-p}^{\prime }\right) \right\} \left\langle 
\mathbf{p}\right\vert \hat{a}(t)\left\vert \mathbf{p}^{\prime }\right\rangle 
\mathrm{d}\mathbf{p}^{\prime }  \label{AS}
\end{equation}%
Equation. (\ref{AR}) can be written in the form 
\begin{equation}
\mathrm{i}\frac{\partial \psi _{\mathbf{p,q}}(t)}{\partial t}=\frac{1}{%
\varepsilon }\left( h\left( t,\mathbf{p}+\varepsilon \mathbf{P}\right)
-h\left( t,\mathbf{p}\right) \right) \psi _{\mathbf{p,q}}(t)\text{,}
\label{AT}
\end{equation}%
where we have noted that $\mathrm{i}\frac{\partial \psi _{\mathbf{p,q}}(t)}{%
\partial \mathbf{q}}=P_{j}\psi _{\mathbf{p,q}}(t)$ in accordance with the
relation $\left[ \hat{a}^{\ast }P_{j}\hat{a},\hat{a}\right] =P_{j}\hat{a}$.
The solution of the linear equation (\ref{AT}) with the initial condition $%
\varphi _{\mathbf{p}}=\psi _{\mathbf{p,}0}(t_{0})$ can be written in the
form (\ref{AP}), as we wish to prove.
\end{proof}
\end{proposition}

If the function $h\left( t,\mathbf{p}\right) $ in the limit $\varepsilon
\longrightarrow 0$ has a $\mathbf{p}$-differentiable limit, then (\ref{AP})
obviously takes the form of the soliton solution (\ref{N}) of the Hartree
equation. In this case, for $\varepsilon \neq 0$, the vector function (\ref%
{AP}) can be called an $\varepsilon $ soliton, or an extremon \cite{3}.

\section{Appendix}

Taking into account equations (\ref{S}) and (\ref{Y}), we write the
representing functions 
\begin{eqnarray}
\left( \gamma \cdot \alpha \right) \left( \rho \right) &=&\exp \left\{ -%
\frac{1}{\varepsilon }\left\langle \rho ,I\right\rangle \right\}
\sum_{n=0}^{\infty }\frac{1}{n!\varepsilon ^{n-1}}\left\langle \rho
^{\otimes n},H^{(n)}\cdot A^{(n)}\right\rangle \text{,} \\
\left\{ \gamma ,\alpha \right\} \left( \rho \right) &=&\exp \left\{ -\frac{1%
}{\varepsilon }\left\langle \rho ,I\right\rangle \right\} \sum_{n=0}^{\infty
}\frac{1}{n!\varepsilon ^{n-1}}\left\langle \rho ^{\otimes n},\left\{
H^{(n)},A^{(n)}\right\} \right\rangle
\end{eqnarray}%
for the products $H^{(n)}\cdot A^{(n)}$, $\left\{ H^{(n)},A^{(n)}\right\} $
in the symbolic form

\begin{equation}
\mathrm{e}^{-\frac{1}{\varepsilon }\left\langle \rho ,I\right\rangle
}\sum_{n=0}^{\infty }\frac{1}{n!\varepsilon ^{n+1}}\left\langle \rho
^{\otimes n},\left( I+\varepsilon \delta _{1}\right) ^{\otimes n}\cdot
\left( I+\varepsilon \delta _{2}\right) ^{\otimes n}\right\rangle \gamma
\left( \rho _{1}\right) \alpha \left( \rho _{2}\right) \mid _{\rho
_{1}=0=\rho _{2}}\text{,}  \notag
\end{equation}%
\begin{equation}
\mathrm{e}^{-\frac{1}{\varepsilon }\left\langle \rho ,I\right\rangle
}\sum_{n=0}^{\infty }\frac{1}{n!\varepsilon ^{n+1}}\left\langle \rho
^{\otimes n},\left\{ \left( I+\varepsilon \delta _{1}\right) ^{\otimes
n},\left( I+\varepsilon \delta _{2}\right) ^{\otimes n}\right\}
\right\rangle \gamma \left( \rho _{1}\right) \alpha \left( \rho _{2}\right)
\mid _{\rho _{1}=0=\rho _{2}}\text{,}  \notag
\end{equation}%
where $\delta _{1}$ and $\delta _{2}$ are the derivatives with respect to $%
\rho _{1}$ and $\rho _{2}$ respectively. Using further Equations. \ref{U}
for the powers $H^{(n)}=\left( I+\varepsilon \delta _{1}\right) ^{\otimes n}$%
, $A^{(n)}=\left( I+\varepsilon \delta _{2}\right) ^{\otimes n}$, we obtain 
\begin{gather}
\left\langle \rho ^{\otimes n},\left( I+\varepsilon \delta _{1}\right)
^{\otimes n}\cdot \left( I+\varepsilon \delta _{2}\right) ^{\otimes
n}\right\rangle =  \label{A3} \\
\left\langle \rho ,\left( I+\varepsilon \delta _{1}\right) \cdot \left(
I+\varepsilon \delta _{2}\right) \right\rangle ^{n}\text{,}  \notag
\end{gather}%
\begin{gather}
\left\langle \rho ^{\otimes n},\left\{ \left( I+\varepsilon \delta
_{1}\right) ^{\otimes n},\left( I+\varepsilon \delta _{2}\right) ^{\otimes
n}\right\} \right\rangle =  \label{A7} \\
n\left\langle \rho ,\left( I+\varepsilon \delta _{1}\right) \cdot \left(
I+\varepsilon \delta _{2}\right) \right\rangle ^{n-1}\left\langle \rho
,\left\{ \varepsilon \delta _{1},\varepsilon \delta _{2}\right\}
\right\rangle .  \notag
\end{gather}%
Substituting this into the preceding formulas, we obtain 
\begin{equation}
\left( \gamma \cdot \alpha \right) \left( \rho \right) =\frac{1}{\varepsilon 
}\mathrm{e}^{\frac{1}{\varepsilon }\left\langle \rho ,\left( I+\varepsilon
\delta _{1}\right) \cdot \left( I+\varepsilon \delta _{2}\right)
-I\right\rangle }\gamma \left( \rho _{1}\right) \alpha \left( \rho
_{2}\right) \mid _{\rho _{1}=0=\rho _{2}}\text{,}  \notag
\end{equation}%
\begin{equation}
\left\{ \gamma ,\alpha \right\} \left( \rho \right) =\mathrm{e}^{\frac{1}{%
\varepsilon }\left\langle \rho ,\left( I+\varepsilon \delta _{1}\right)
\cdot \left( I+\varepsilon \delta _{2}\right) -I\right\rangle }\left\langle
\rho ,\left\{ \delta _{1}\gamma \left( \rho _{1}\right) ,\delta _{2}\alpha
\left( \rho _{2}\right) \right\} \right\rangle \mid _{\rho _{1}=0=\rho _{2}}%
\text{.}  \notag
\end{equation}%
Multiplying the exponentials and noting that $\mathrm{e}^{\left\langle \rho
,\delta ^{\prime }\right\rangle }\beta (\rho ^{\prime })\mid _{\rho ^{\prime
}=0}=\beta (\rho )$, we obtain 
\begin{equation}
\left( \gamma \cdot \alpha \right) \left( \rho \right) =\frac{1}{\varepsilon 
}\mathrm{e}^{\varepsilon \left\langle \rho ,\delta _{1}\cdot \delta
_{2}\right\rangle }\gamma \left( \rho _{1}\right) \alpha \left( \rho
_{2}\right) \mid _{\rho _{1}=\rho =\rho _{2}}  \label{A5}
\end{equation}%
\begin{equation}
\left\{ \gamma ,\alpha \right\} \left( \rho \right) =\mathrm{e}^{\varepsilon
\left\langle \rho ,\delta _{1}\cdot \delta _{2}\right\rangle }\left\langle
\rho ,\left\{ \delta _{1}\gamma \left( \rho _{1}\right) ,\delta _{2}\alpha
\left( \rho _{2}\right) \right\} \right\rangle \mid _{\rho _{1}=\rho =\rho
_{2}}\text{.}  \label{A6}
\end{equation}%
Expanding the exponential (\ref{A5}) in a series in $\varepsilon $, we
obtain the expansion (\ref{AC}). The first formula (\ref{AD}), from which
Proposition (\ref{P1}) follows, is obtained by applying the second formula (%
\ref{A5}) to (\ref{A6}), and the expansion (\ref{AD}) in powers of $%
\varepsilon $ can be readily obtained by using formulas of the type (\ref{A3}%
) and (\ref{A7}) in reverse order in the exponential (\ref{A6}).

\end{document}